\documentclass[pre,twocolumn,showpacs,preprintnumbers,amsmath,amssymb]{revtex4}
\def\beneq{\begin{equation}}
\def\eneq{\end{equation}}
\def\bea{\begin{eqnarray}}
\def\eea{\end{eqnarray}}
\usepackage{subfigure}
\usepackage{graphicx}
\usepackage{dcolumn}
\usepackage{bm}
\begin{document}


\title{Plug flow and the breakdown of Bagnold scaling in cohesive granular flows}
\author{Robert Brewster,$^{1}$ Gary S.\ Grest,${^2}$ James W.\ Landry,$^{3}$ and Alex J.\ Levine$^{1}$}
\affiliation{$^{1}$Department of Chemistry and Biochemistry, UCLA, Los Angeles, CA 90095-1596\\
$^{2}$Sandia National Laboratories, Albuquerque NM 87185 \\
$^{3}$ BAE Systems Burlington, MA 01803}
 
\date{\today}

\begin{abstract}
Cohesive granular media flowing down an inclined plane are
studied by discrete element simulations. Previous work on
{\em cohesionless} granular media demonstrated that within
the steady flow regime where gravitational energy is
balanced by dissipation arising from intergrain forces, the
velocity profile in the flow direction scales with depth in
a manner consistent with the predictions of Bagnold. Here
we demonstrate that this Bagnold scaling does not hold for
the analogous steady-flows in cohesive granular media. We
develop a generalization of the Bagnold constitutive
relation to account for our observation and  speculate as
to the underlying physical mechanisms responsible for the
different constitutive laws for cohesive and noncohesive
granular media.
\end{abstract}
 \maketitle

\section{Introduction}

One of the central questions in the study of granular flows
is how to determine the relationship between the
microphysics of grain interactions and the
collective or macroscopic flow properties of the system.
Elucidating these flow properties is of fundamental
importance to a variety of fields ranging from civil
engineering to geophysics \cite{Pouliquen:02}. Moreover,
this question remains at the forefront of many-body physics
as its solution appears to demand entirely new concepts
that are applicable to systems driven far from equilibrium.

In this article we explore the velocity field in steadily
flowing granular media in the chute geometry -- an inclined
plane having a rough surface at its base and a free
surface at the top. Such flows have
been previously characterized both
experimentally \cite{Suzuki:71,Augenstein:78,Drake:90,Ancey:96,
Pouliquen:99,Azanza:99,Ancey:02,Pouliquen:04} and via
simulation
\cite{Walton:93,Zheng:96,Dippel:99,Hanes:00,Silbert:01,Silbert:02,Silbert:03}.
In this work we address, through large-scale discrete
element simulations, the change in the flow field in the
material as a function of interparticle adhesion. We
present a modified constitutive law relating shear stress
within the bulk to the local gradients in the velocity
field and discuss the implications of our proposed
continuum description of the material. In addition we
observe the formation of a plug flow regime in the cohesive
slab extending from the free surface at the top of the slab
into the bulk. The depth of this plug flow regime can be
calculated from a comparison of the yield stress of the
cohesive material to depth-dependent shear stress due to
the weight of the slab.

The study of cohesive granular media is necessary for the
application of granular physics to the behavior of such
materials in their geophysical context, {\em i.e.} the
dynamics of landslides, snow avalanches, and even the lunar
regolith. In the former examples, a small aqueous wetting
layer on the particles forms menisci at intergrain contacts
to produce an adhesive force. In the latter case,
high-vacuum conditions facilitate close contacts between
particles leading to interparticle adhesion due to van der
Waals interactions.   The study of cohesive granular
materials also helps to elucidate fundamental questions
concerning the physical description of this state of
matter. One aspect of granular materials that
differentiates them from random elastic solids or liquids
is that cohesionless granular system cannot support tensile
stresses.  By observing the quantitative effect of known
cohesive forces upon granular flows, one should be able to
bridge between the better understood elastic solids and/or
viscous liquids and granular materials, which, by
comparison, remain rather mysterious. The change in the stress-strain relation due to internal cohesion also
provides insight into the rheology of cohesionless granular matter.  We discuss this further below.

In the study of granular mechanics it is clearly desirable
to develop a continuum description of flow since
calculating the detailed dynamics of numerous intergrain
collisions rapidly becomes intractable with increasing
numbers of such particles. In addition one expects, based
on experience with the continuum mechanics of solids,
liquids, and gases, to be able to develop a  set of
relations between macroscopic averaged observable such as
the velocity field $v_\alpha(\mathbf{x})$, mass density
$\rho(\mathbf{x})$, and stress tensor $\sigma_{\alpha
\beta}(\mathbf{x})$ within the steadily-flowing granular
material where details of the grain interactions enter
through a small set of parameters
\cite{Haff:83,Savage:83,Savage:98}. Such a constitutive
relation between the stress state in the material and its
rate of deformation taken together with momentum
conservation completely determines the flow properties of
the granular system assuming the boundary conditions on the
flow are sufficiently well-known.  There are now a number
of such hydrodynamic descriptions of granular flow which
seek to derive such constitutive equations from a more
microscopic theory \cite{Jenkins:83,Haff:83}.
There are a variety of such models that rely on arguments
based on effective
viscosities \cite{Losert:00,Bocquet:01,Bocquet:02},
transient force-chains\cite{Mills:00} or
a superposition of a rate-dependent contribution arising from 
collisional interactions and a rate-independent part related to 
enduring frictional contacts among the grains\cite{Louge:03}.

A well-known proposal by Bagnold\cite{Bagnold:54} for such
a constitutive relation for granular flows is that the
shear stress in the flow is proportional to the
instantaneous square of the rate of strain tensor and that
the density is constant throughout the material. Taking (as
we do throughout this work) a coordinate system where the
$x$-axis is directed in the flow direction and the $z$-axis
as the upward normal to the inclined plane, the Bagnold
relation is
\begin{equation}
\label{Bagnold} \sigma_{xz} = \kappa \left( \partial_z v_x
\right)^2,
\end{equation}
where $\kappa$ is a constant having the dimensions of
$M/L$. One can reduce the supporting argument for
Eq.~\ref{Bagnold} to dimensional analysis; in the flowing
granular system the one energy density available is the
kinetic energy of the grains themselves $\sim \rho v^2$. As
long as the local mass density remains constant, Galilean
invariance requires the shear stress to be proportional to
the square of the velocity difference across the sample.
The further assumption that the relationship between stress
and the rate of strain is local produces Eq.~\ref{Bagnold}.
A simple heuristic argument to derive the scaling relation
shown in Eq.~\ref{Bagnold} and attributed to Bagnold is as
follows: the transport in the gradient direction
($\hat{z}$) of the component of momentum parallel to the
velocity direction ($\hat{x}$) occurs only through
collisions between grains. The rate of those collisions
depends on the average velocity difference between grains
at different locations along the gradient direction $\sim a
\partial_z v_x$ where $a$ is the grain radius. The momentum
transfer per collision also scales linearly with the
velocity difference leads to $\sigma_{xz} \sim m (a
\partial_z v_x)^2$ where $m$ is the mass of a single grain.

It is instructive to compare the above argument to the
standard expression for the viscosity of a gas as
determined by kinetic theory. In that case the rate of
intermolecular collisions is controlled by the root mean
square velocity of the gas molecules that is fixed by the
equipartition theorem. Thus the viscosity of the gas is
proportional to the square root of temperature. The
granular system is effectively at zero temperature so that
the random particle velocities are driven themselves by the
macroscopic velocity scale making the effective granular
viscosity as extracted from Eq.~\ref{Bagnold} depend on
shear rate.

In the remainder of this article we first discuss the
simulation method in section \ref{method} before turning to
a discussion of our numerical results and calculations in
section \ref{results}. In that section we discuss the
effect of intergrain adhesion on the validity of the
Bagnold constitutive law.  We then conclude in section
\ref{summary}.

\section{Simulation Method}
\label{method}

\begin{figure}[tpb]
\begin{center}
\includegraphics[width=2cm]{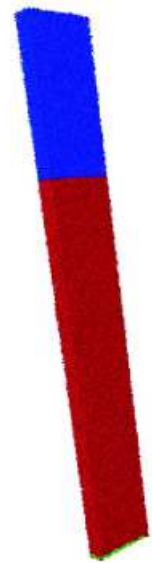}
\caption{(color online) A representative example of the
flowing granular slab of height $H=300$. In this cohesive
granular slab ($A=0.8$) the system separates into a
solid-like plug (blue)  sliding upon a flowing granular bed
(red). The rough surface of inclined plane is shown in
green.} \label{slab-picture}
\end{center}
\end{figure}

We performed discrete element simulations on a
three-dimensional system of N monodisperse particles of
mass $m$ and diameter $d$ on a rough base tilted an angle
$\theta$ with respect to gravity.  Our simulations volume
is a rectangular box with periodic boundary conditions in
both the $x$ and $y$ directions, a rough base and an open top.
The rough base is created by taking a slice through a
previously random close packed state of particles with the same
diameter $d$.  We define the z-axis to be normal to the
base and the $x$-axis as the direction of flow. We
study a system of length $40d$ and $10d$ in the $x$ and $y$
directions respectively. We studied three granular slabs of
differing heights $H$, containing 
$~42000$, $83000$, and $125000$ particles respectively. 
We refer to these three systems by their approximate height, namely $H=100d$, $200d$, and $300d$. We show an example of a cohesive granular material in the steadily flowing state in Fig.~\ref{slab-picture} for $H=300d$.

We employ a modified version of the model developed by
Cundall and Strack \cite{Cundall:79} to model cohesionless  particulates.
The model uses Hookean contacts
to model grain to grain interactions.  We add 
a cohesive force between particles to simulate damp
granular media.  Two contacting spheres $i$ and $j$ with
positions $\mathbf{r}_{i}$ and $\mathbf{r}_{j}$ are
separated by
$\mathbf{r}_{ij}=\mathbf{r}_{i}-\mathbf{r}_{j}$, the
compression is then $\delta_{ij} =d-|\mathbf{r}_{ij}|$.
The relative velocity is
$\mathbf{v}_{ij}=\mathbf{v}_{i}-\mathbf{v}_{j}$ which can
be separated into normal and tangential components
\begin{eqnarray}
\mathbf{v}_{n_{ij}}&=&(\mathbf{v}_{ij}\cdot\mathbf{n}_{ij})\mathbf{n}_{ij} \\
\mathbf{v}_{t_{ij}}&=&\mathbf{v}_{ij}-\mathbf{v}_{n_{ij}}-
\frac{1}{2}(\mathbf{\omega}_{i}+\mathbf{\omega}_{j})\times\mathbf{r}_{ij}.
\end{eqnarray}
The normal and tangential forces acting on particle $i$ due
to particle $j$ can then be written as
\begin{eqnarray}
\label{normal-force}
\mathbf{F}_{n_{ij}}&=&(k_{n}\delta_{ij}\mathbf{n}_{ij}-
\frac{m}{2}\gamma_{n}\mathbf{v}_{n_{ij}}) + \mathbf{F}^{\rm c}_{ij}\left( \delta_{ij} \right) \\
\label{tangential-force}
\mathbf{F}_{t_{ij}}&=&(-k_{t}\mathbf{u}_{t_{ij}}-
\frac{m}{2}\gamma_{t}\mathbf{v}_{t_{ij}}),
\end{eqnarray}
where $k_{n}, k_t$ and $\gamma_{n}, \gamma_{t}$ are the
elastic and viscoelastic constants for interparticle motion
along ($n$) and normal to the line of centers ($t$).
$\mathbf{u}_{t_{ij}}$ is the elastic tangential
displacement that is set equal to zero at the initiation of
contact and is truncated to satisfy the Coulomb yield
criterion,
$|\mathbf{F}_{t_{ij}}|<\mu|\mathbf{F}_{n_{ij}}|$. The
additional normal force $\mathbf{F}^{\rm c}_{ij}\left(
\delta_{ij} \right)$, is the cohesive normal force between
particles $i$ and $j$. This normal force is derived from an
effective cohesive potential acting on particle $i$ due to
particle $j$. We simply choose a gaussian well centered
around particle $j$
\begin{equation}
\label{cohesion-pot}
U^c_{ij}=-Ae^{\frac{-\delta_{ij}^{2}}{\ell ^{2}}},
\end{equation}
where $A$ is the strength of the attraction (in units of mgd)
and $\ell$ is the width of the well.  The force is then
$\mathbf{F}^c_{ij}=-\mathbf{\nabla}U^c_{ij}.$  We focus on
$0.0\leq A\leq 1.0$, where  $A=1.0$ corresponds to a force
capable of supporting $~85$ particles under gravity. While
this form of the cohesive potential cannot be justified in
terms of details of either van der Waals or wetting-layer mediated interparticle interactions, we choose this
form so that we can vary the strength of the interparticle
cohesion via the well depth $A$ while maintaining the
short-range nature of the interaction, which is controlled
by $\ell$.  The force captures the essential features of
more physical adhesion scenarios and we  expect that the
effects of interaction cohesion as modelled here, to
accurately reflect the behavior of experimentally
accessible cohesive granular media.

Most of the simulations were run with 
$k_n=2\times 10^{5}\frac{mg}{d}$, $k_t=\frac{2}{7}k_n$,
$\gamma_n=50\tau^{-1}$, and $\gamma_t=0$, where $\tau = \sqrt{d/g}$.  
For the case of  Hookean springs used here,  these parameters give
a coefficent of restitution $e_n=0.88$ force normal collisions.
The normal spring constant
$k_n$ is large enough to minimize the overlap between particles
but small enough to be computationally efficient. Previous
simulations for cohesionless grains  \cite{Silbert:01} found that
this value of $k_n$ 
provides results similiar to those for larger values of $k_n$.
However we do find a subtle
dependence of the flow rheology upon the value of $k_n$
which we explore later in this article. 
For the coefficent of
friction, we use $\mu =0.5$, which is typical for the types of
materials we are modeling. 
The normal damping
term operates only when the particles are in actual contact
so that, in a narrow window of interparticle separations $d
< r < d + \ell$ these particles will experience a cohesive
force without the velocity dependent damping. Interpreted
in terms of action of interparticle wetting layers, this
suggests that we discount dissipative hydrodynamics in the
wetting layer over microscopic plastic deformation in the
particle themselves. We expect that such hydrodynamic
effects modify the effective value of $\gamma_n$ as well as
the interparticle separation over which it operates, but we
have not attempted to model this particular cohesion
scenario in detail.

In a gravitational field $\mathbf{g}$, the total force on a
particle from
Eqs.~\ref{normal-force}, \ref{tangential-force}, and
\ref{cohesion-pot}:
\begin{equation}
\label{total-force}
\mathbf{F}^{tot}_{i}=m_{i}\mathbf{g}+\sum_{j}
\left( \mathbf{F}_{n_{ij}}+ \mathbf{F}_{t_{ij}} +
\mathbf{F}^c_{ij} \right),
\end{equation}
where the sum is over neighboring particles.

The stress tensor within a volume $V$ is computed by
summing over both the contact and kinetic terms of each
particle within that volume
\begin{equation}
{\sigma}_{\alpha\beta}=\frac{1}{V}\sum_{i}[\sum_{i\neq
j}\frac{r^{\alpha}_{ij}F^{\beta}_{ij}}{2}+m_{i}(v^{\alpha}_{i}-
\overline{v^{\alpha}})(v^{\beta}_{i}-\overline{v^{\beta}})],
\end{equation}
where
$\mathbf{F}^{\beta}_{ij}=\mathbf{F}^{\beta}_{t_{ij}}+\mathbf{F}^{\beta}_{n_{ij}}
+ \mathbf{F}^{{\rm c}\beta}_{ij}$ and $\overline{v}$ is the
time-averaged velocity of the particles in $V$.

The timestep for the integration of the
equations of motion is $\delta t = 10^{-4} \tau$.
After equilibration the systems were typically run
between $1-5\times 10^7$ time steps. Steady state was
determined using the total kinetic
energy of the system as a criterion for 
suitable equilibration of each sample.

\section{Results}
\label{results}

\begin{figure}[htbp]
\includegraphics[width=6cm]{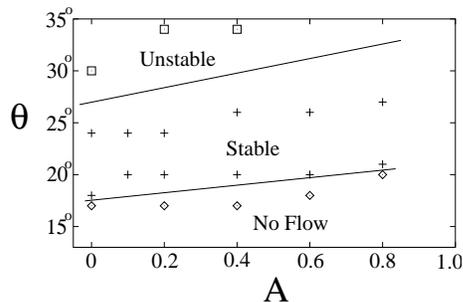}
\caption{Phase diagram for chute flow of a fixed height
granular slab. The diagram is spanned by the tilt angle
$\theta$ and strength of cohesion $A$.  There exist three
well defined regions corresponding to no flow, stable flow
and unstable flow.  Lines are drawn to seperate the regions.} \label{phase}
\end{figure}

In chute flow there are three qualitatively distinct
regimes determined by the height of the slab $H$, the
inclination angle $\theta$ of the lower surface with
respect to the direction of gravity
\cite{Silbert:01,Pouliquen:99}, and the intergrain cohesive
stresses determined in our model by $A$ -- see
Eq.~\ref{cohesion-pot}.  For small enough inclination
angles ({\em i.e.} below the maximum critical angle) or
alternatively for short and/or cohesive slabs, the granular
heap is stationary. Below the angle of repose for a given
slab height and cohesive energy, transiently flowing states
of the slab dissipate energy more rapidly than the input
of gravitational potential energy so that the slab stops.
We refer to this as the {\rm no flow} regime. At much
larger angles, on the other hand, the energy dissipation
with the slab is less than the gravitational energy input
so that the slab continuously accelerates down the plane.
We refer to the parameter space exhibiting this behavior as
the {\rm unstable regime}. At angles intermediate between
these two regimes for a given cohesive energy and slab
height, we observe steady-state flows. In this work we
concentrate primarily on such steady-state behavior in
slabs of the largest heigth $H=300d$. 
The phase diagram 
spanned by the cohesion parameter $A$ and the inclination
angle of the slab $\theta$
for this system shown in Fig.~\ref{phase}.
In this work we investigate the
steady-state flowing regime after transient behavior
associated with the initiation of flow have decayed. The
onset of such steady-state behavior is determined by
observing the total kinetic energy of the system.

\subsection{Plug-flow}

\begin{figure}
\begin{center}
\includegraphics[height=1.4in]{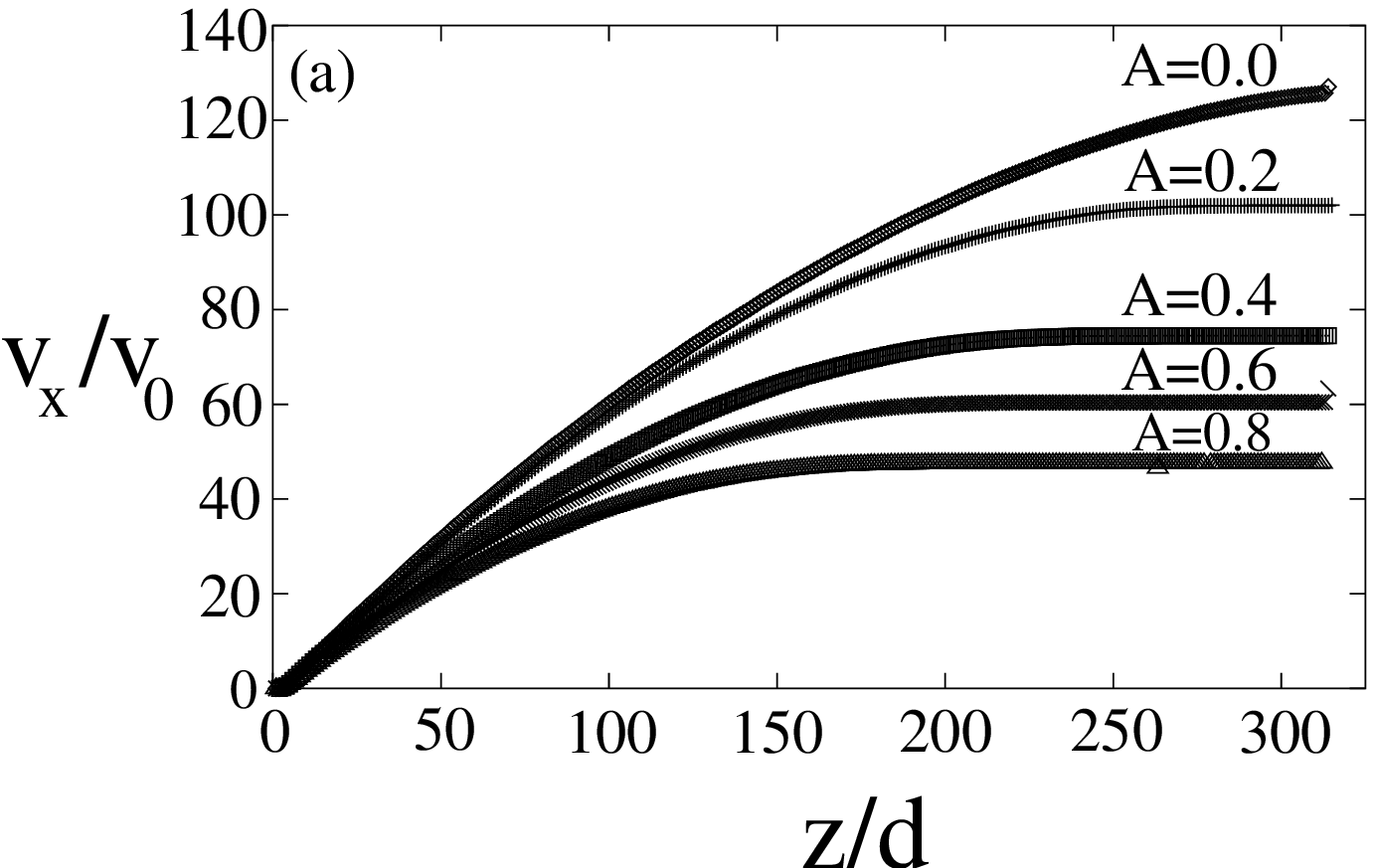}
\includegraphics[height=1.4in]{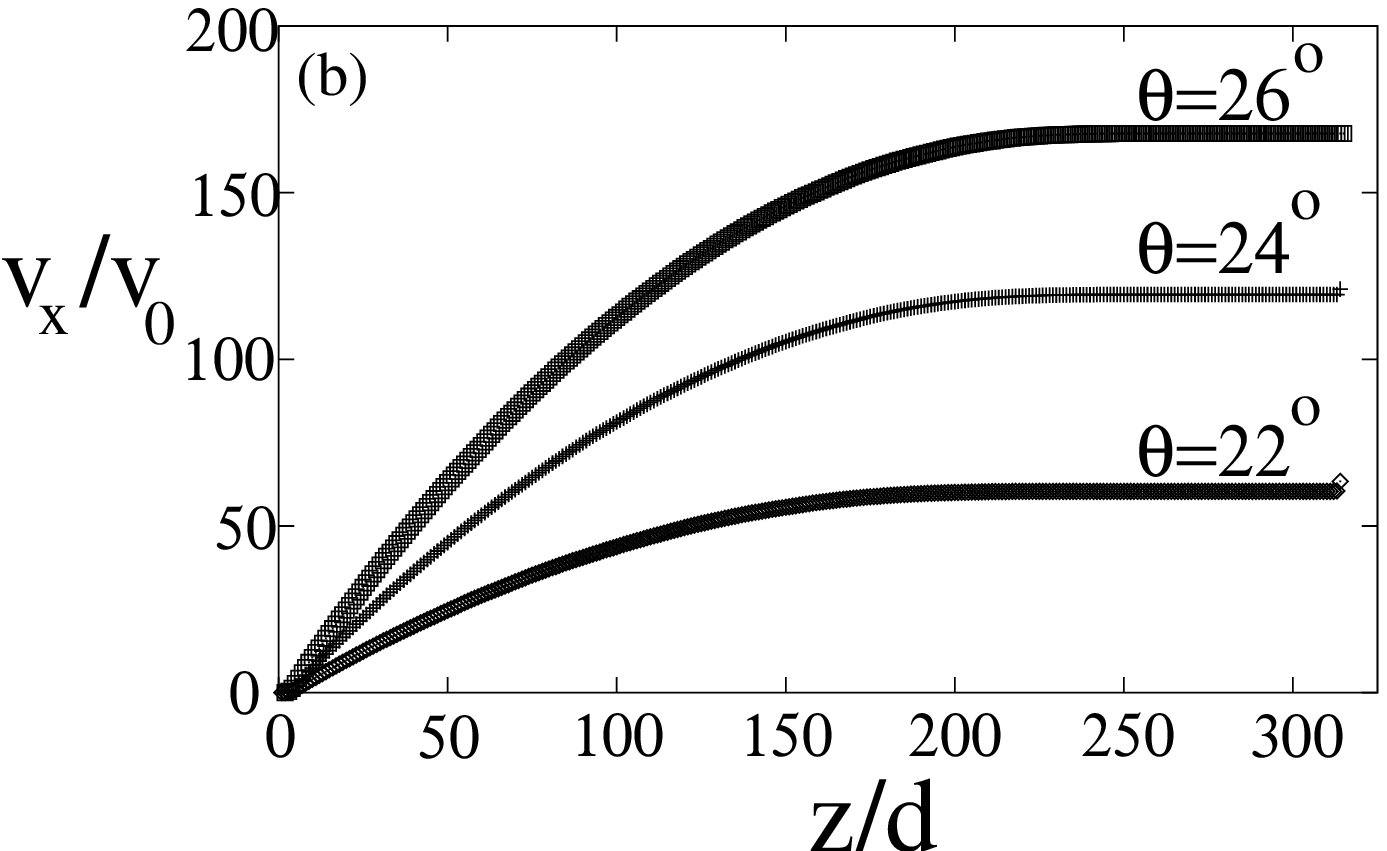}
\includegraphics[height=1.4in]{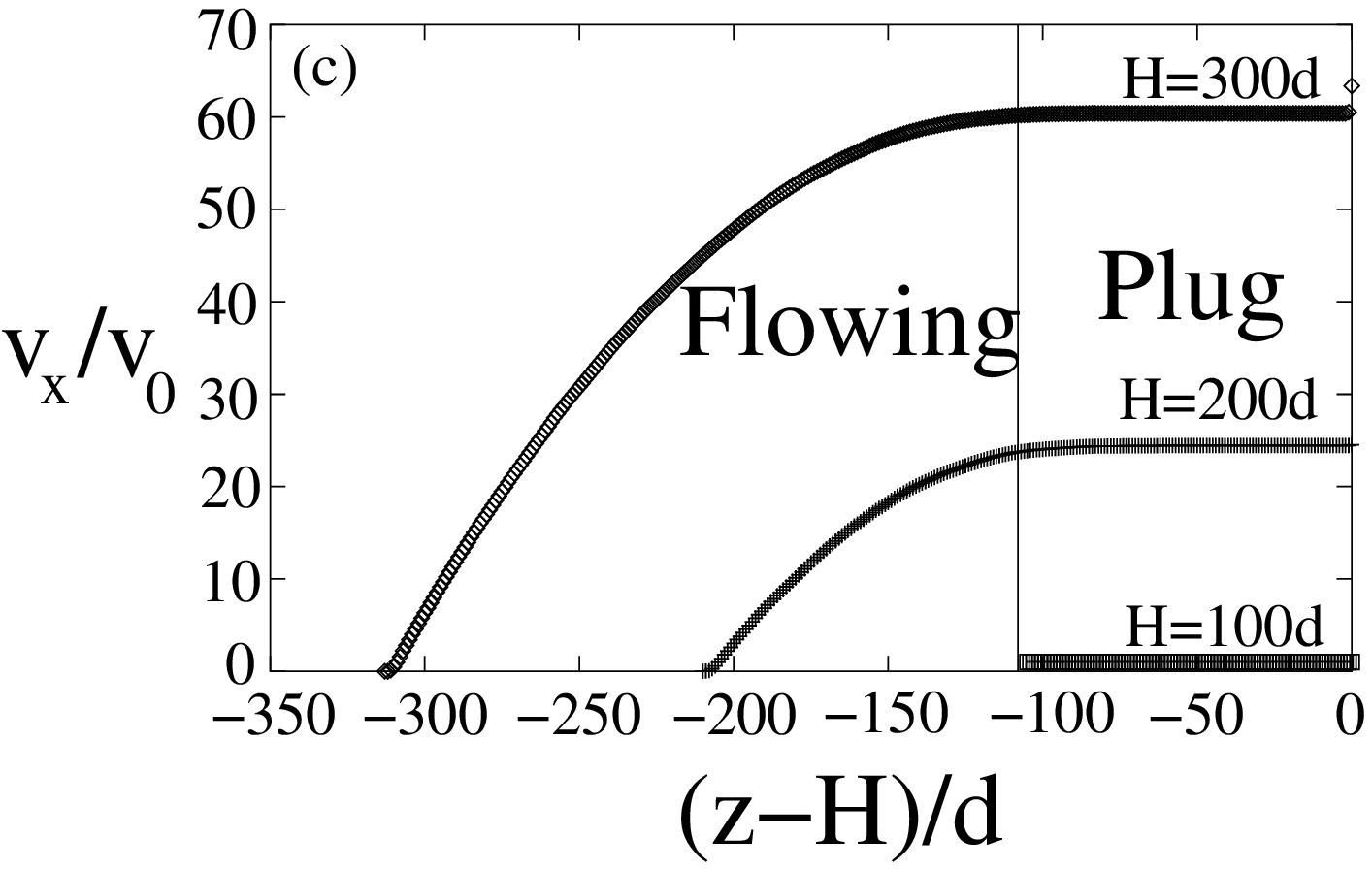}
\caption{Velocity profiles as a function of $z$.  In all three figures the velocity is measured in units of $v_0 = \sqrt{g d}$. In (a)
profiles are shown for different values of $A$ with $\theta
=22^{o}$, $H=300 d$. In (b)  velocity profiles for
different values of the tilt angle $\theta$ with $A=0.6$,
$H=300 d$. And (c) the velocity profile as a function of
the depth from the free surface for differing slab heights
$H$ with $A =0.6$, $\theta =22^{o}$. For $H=100 d$,  only a
stopped plug is observed as $H$ is similar to the size of
the plug, w. In each profile two distinct regions are
visible, a flowing state at depth and a plug flow of
varying thicknesses at the free surface. For the
noncohesive case $A=0.0$ the plug size vanishes.}
\label{profiles}
\end{center}
\end{figure}

In Fig.~\ref{profiles}a we plot the velocity in the flow
direction vs. height from the bottom of the slab ($z$) for
a range of values of the cohesive energy $A$. In each case
the inclination angle is $\theta =22^{o}$. One of the
central results of this work is immediately evident in
these figures. For nonzero values of the intergrain
cohesive energy $A$, the velocity plateaus at a finite
fraction of the height of the slab. The granular material
has, in effect, phase separated into a liquid-like flowing
region and a solid-like region that does not admit a
non-zero rate of the shear strain. We refer to this
behavior near the free surface as plug flow and the
solid-like region as the plug. It is clear from this figure
that the thickness of the plug depends on the cohesive energy
$A$.  Figure \ref{profiles}b shows the dependence on the plug
thickness at fixed $A$ on the inclination angle $\theta$.
Finally, Fig.~\ref{profiles}c demonstrates that the
thickness of the plug is independent of the total height of
the slab as long as the material remains in the flowing
state for that slab height. Therein the granular velocity
down the chute for three slabs of heights $H=100d$, $200d$,
and $300 d$ is plotted as a function of the depth from the
free surface for an inclination angle of $\theta = 22^{o}$
and $A=0.8$.

\begin{figure}
\begin{center}
\includegraphics[height=1.4in]{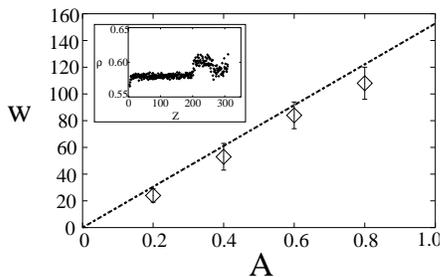}
\caption{The thickness of the plug w vs.\ the cohesive
energy $A$ for a fixed angle of inclination, $\theta =
22^{o}$, and pile height, H=300d.  The dashed line shows the plug width
prediction from Eq.~\ref{plug-thickness}. The uncertainties
in the plug thickness are determined from the width of the
density jump at the lower boundary of the plug. Note that the plug boundary becomes more diffuse for smaller values of $A$.}
\label{fig:plug-thickness}
\end{center}
\end{figure}

The development of the plug in steady state is signalled
not only by the appearance of the velocity plateau, but
also by the development of the localized jump in the
particle volume fraction with height. The flowing material
below due to shear dilatancy is generically at a lower
volume fraction, while the plug is denser. Typically this
volume fraction change is on the order of $4 \%$ and occurs
over a height range of $4 - 10$ particle diameters. We use
this abrupt change in the particle volume fraction with
height to more precisely determine the interface between
the flowing and plug regimes. The inset of
Fig.~\ref{fig:plug-thickness} shows a typical example of
this volume fraction change at the plug boundary.

To interpret these observations, we propose that the
cohesive granular material can support a finite yield
stress $\sigma_c$ before flowing. Since shear flow requires dilatancy, the critical yield stress should be equal to the typical cohesive stress in the pile.  We estimate the maximal value of this cohesive stress as follows. We assume all contacts provide some average adhesive force and then calculate the mean number of such interparticle contacts per unit cross-sectional area, $n_c$, in terms of the volume fraction of the particles, $\phi$. The yield stress is then estimated as the product of the average adhesive force per contact and $n_c$. To determine the average adhesive force per contact, $\bar{f}$, we assume that the interparticle separations  at each contact are randomly distributed within the attractive potential well between $d$ and $d+ \ell$. Using this assumption we find that the mean magnitude of the adhesive force is given by
\begin{equation}
\label{average-force}
\bar{f} = \frac{ \left| U(0) - U(\ell) \right| }{\ell},
\end{equation}
in terms of the potential $U$ defined in Eq.~\ref{cohesion-pot}.
From Eq.~\ref{average-force} and $n_c = 3 \phi/(2 \pi a^2)$ we find that the yield stress is  
\begin{equation}
\label{yield-stress} 
\sigma_{c}=\frac{3 \phi \bar{f}}{2 \pi a^2}.
\end{equation}
We determine the thickness of the plug w by equating
the shear stress in the slab due to the weight of the
overlying material to this yield stress.  We find the
thickness of the plug is
\begin{equation}
\label{plug-thickness} 
{\rm w} =\frac{A}{\ell \sin \theta } \left[ \frac{e-1}{e} \right],
\end{equation}
where $e$ is the base of the natural logarithm. 
To test this proposal we plot in
Fig.~\ref{fig:plug-thickness} the calculated plug from
Eq.~\ref{plug-thickness} with that measured by particle
density change at the plug boundary.

Over the parameter range tested, the calculated thickness
of the plug is in reasonable agreement with the data. Clearly the
linear dependence of the plug size on the cohesive energy
is supported by the data. Precise tests of the $1 / \sin \theta$ dependence of the plug thickness, on the other hand, are difficult due to the limited range of available angles for which the slab exhibits steady-state flow. Our calculation of the slope of the theoretical curve must be treated as an upper bound as it is based on the assumption that all interparticle contacts generate some attractive force. Due to jamming in disordered sphere packings, it is reasonable to suppose that at least some contacts have interparticle spacings less than $d$ and are therefore repulsive in nature. Such contacts between jammed spheres reduce the effective yield stress.

\subsection{Testing Bagnold-scaling}

From the Bagnold conjecture, Eq.~\ref{Bagnold}, for the
constitutive relation in the flowing granular state, one
may immediately determine the flow profile down the chute.
The resulting velocity profile takes the form:
\begin{equation}
\label{Bag-vel} v_{x}(z)=2 \sqrt{ \frac{H^3\rho g \sin
\theta}{9 \kappa
}}\left[1-\left(\frac{H-z}{H}\right)^{3/2}\right].
\end{equation}
In Appendix \ref{Bag-derivation} we derive this velocity
profile for the case of a free surface at the top of the
flowing medium, {\em i.e.} no plug as well as the analogous
velocity profile for the case of a finite shear stress
applied to the top of the flowing state due to the weight
of the plug.

\begin{figure}
\includegraphics[width=7cm]{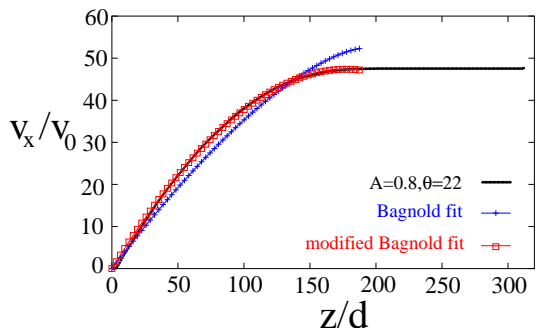}
\caption{(color online) Velocity profile for $A=0.8$, $\theta=22^{o}$ showing the difference in the standard Bagnold fit (blue $+$) and the modified Bagnold fit (red $\square$).} 
\label{a8ang22}
\end{figure}

In previous research on cohesionless  granular chute flow
($A=0.0$), the down-chute velocity
versus height $v_x(z)$ has been reported as being
consistent with the Bagnold power-law form. We also find
such apparently reasonable agreement -- see the uppermost
curve in Fig.~\ref{profiles}a.  In the presence of cohesion
($A>0$), however, the expected Bagnold velocity profile
fits the data poorly. We interpret this failure of the
Bagnold hypothesis as demonstrating a new mode for
vertically transporting ($z$) down-chute momentum ($p_x$)
due to the presence of long-lived contacts in the material.
These long-lived adhesive contacts in the material create
clusters spanning different streamlines in the flow as has
been proposed by Erta\c{s} and Halsey \cite{Ertas:02}. In
the presence of shearing flow, these forces acting along
these clusters of particles transmit momentum  $p_x$
proportional to the shear rate $\dot{\gamma} = \partial
v_{x}/\partial z$. Thus, for cohesive granular materials the
Bagnold relation can be generalized to a form that is a sum
of terms that are linear and quadratic in the shear rate.
We propose the modified Bagnold relation,
\begin{equation}
\sigma_{xz} - \sigma_c =\kappa (\frac{\partial
v_{x}}{\partial z})^2+\beta (\frac{\partial
v_{x}}{\partial z}).
\end{equation}
The second of these terms represents the new mode of
momentum transport made possible through the long-lived
contact networks in the material while the first term
arises from the short time scale collisions originally
considered by Bagnold \cite{Bagnold:54}.  
The constant stress term on the LHS
of the above equation is the finite yield stress of the
cohesive material. The above relation applies only in the
flowing states, {\em i.e.} for values of the shear stress
greater than the yield stress $\sigma_c$.  The constant
$\kappa$ is the
Bagnold parameter introduced earlier in Eq. 1, while the second
constant $\beta$, having dimensions of a viscosity measures
the relative importance of the long-lived contacts to the
transient collisions. Clearly, this modified constitutive
relation leads to a new velocity profile $v_{x}(z)$ for the
material below the plug. In Fig. \ref{a8ang22} we plot two
best fits to the velocity profile in the flowing state.  We
have chosen data corresponding to $A=0.8,$ $\theta=22^{o}$
that shows a  typical example of the flow profile in the
strongly cohesive limit.  The curve ($+$) is the fit of
the data to the Bagnold velocity profile where we have used
the least-squares method. Note that one cannot
simultaneously fit this initial slope of the velocity
profile and match the curvature of the data.
The modified Bagnold relation gives the best fit to the data.

Simply demonstrating a better fit with the modified Bagnold
form is not, in itself, conclusive evidence of the
breakdown of Bagnold scaling as one expects a better fit
from a model with an extra adjustable parameter.  To better
justify our conjecture we first examine the
dimensionless ratio of the viscous stress proportional to
$\beta \dot{\gamma}$ to the Bagnold stress, $\kappa
\dot{\gamma}^2$:
\begin{equation}
\Omega=\frac{\beta}{\kappa \dot{\gamma}}.
\end{equation}
For the Bagnold hypothesis, $\Omega$ vanishes. In
essence we can determine the extent of the breakdown of
such Bagnold scaling by extracting the fit parameters
$\kappa,\beta$, as well as the shear rate from velocity
profiles as shown in Fig.~\ref{a8ang22} in order to compute
the value of  $\Omega$ from our numerical data. This ratio
can be computed for a variety of samples having different
tilt angle and different cohesive energies. In addition
the ratio can be computed at different heights within the
flowing layer of a given sample.

\begin{figure}
\includegraphics[width=6cm]{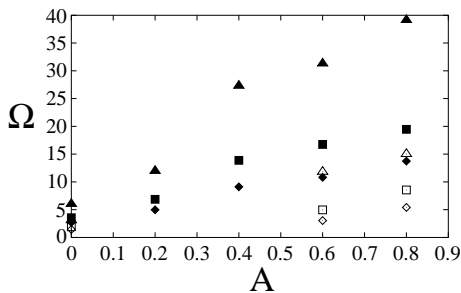}
\caption{ $\Omega$ vs $A$ for $\theta=22^{o}$ (closed symbols), $24^{o}$ (open symbols) at
three different heights in the flowing slab ($\diamond$ for $z=H/4$, $\square$ for $z=H/2$, and $\triangle$ for $z=3H/4$).  $\Omega$ is a
measure of the importance of the linear term in
$\sigma_{xz},$ as $A$ increases so does the importance of
this linear term} \label{bvsa}
\end{figure}

In Fig.~\ref{bvsa}, we show a plot of  
$\Omega$ vs.\ $A$.
Examination of this figure reveals three
trends. First, we note that for any height $z$ within the
flowing part of the slab and for any tilt angle $\theta$,
$\Omega$ increases with increasing cohesive energy $A$.
The deviation for a purely Bagnold constitutive
law as measured by $\Omega$ increases with height in the
slab for all measured angles and cohesive energies. 
For a given height in the slab and a given cohesive energy
one finds that $\Omega$ decreases within increasing tilt
angle $\theta$.

The increase of $\Omega$ with $A$ for all heights and
angles strongly suggests that the internal cohesion is
primarily responsible for the breakdown of the Bagnold
scaling. That the magnitude of the discrepancy between
the observed flows and those derived from the Bagnold
hypothesis depends on the height within the slab ($\Omega$
vs.\ z at fixed $A$ and $\theta$) is consistent with our
hypothesis that long-lived contacts cause this discrepancy.
Clearly as one approaches the plug the intergrain contacts
have extremely long life-times and the constitutive
relation for the material most greatly deviates from the
Bagnold form. It is reasonable to suppose that in the
region directly below the plug where the gravitational
shear stress is only slightly greater than the yield stress
of the material that there would be the greatest density of
long-lived contacts in the flowing state. Thus, one would
expect the greatest deviation from Bagnold scaling near the
plug. The data clearly support this; for all tilt angles and cohesive energies $\Omega$ increases with height in the slab. Finally, as the tilt angle is increased at constant
cohesive energy, the total kinetic energy of the system is
increased. The grains, having higher typical velocities
will now have fewer long-lived contacts and transient
forces associated with brief intergrain collisions will
become the more dominant momentum-transfer process in the
material. Thus the effective constituent law will appear to
be closer to the Bagnold form.

\begin{figure}
\includegraphics[width=6cm]{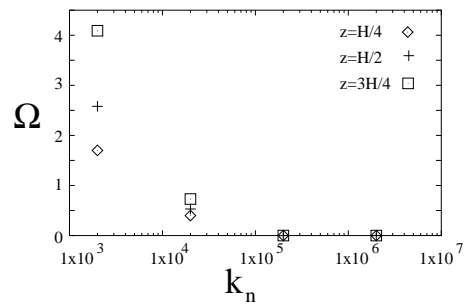}
\caption{$\Omega$ vs $k_n$ for cohesionless granular flows
for $H=100d$ and $\theta=22^o$.
Note that the Bagnold constitutive law ($\Omega = 0$)
desribes the data better as the stiffness of the
particles increases.} \label{kn}
\end{figure}

It is apparent from Fig.~\ref{bvsa} that even in the limit
of no cohesion there remains a significant deviation from
the original Bagnold constitutive law. This residual
discrepancy can be well accounted for by slight
interpenetratibility of the particles in the simulation.
The Bagnold constitutive relation can hold exactly only in
the limit of hard sphere particles \cite{Lois:05}. To
test this point we have examined cohesionless
particles of varying stiffness $k_n$ for $H=100d$.  
The data in Fig.~\ref{kn} shows that the
residual deviations from Bagnold behavior of our
cohesionless model granular material can be attributed to
the finite stiffness of the constituent particles.

\begin{figure}
\begin{center}
\includegraphics[height=2in]{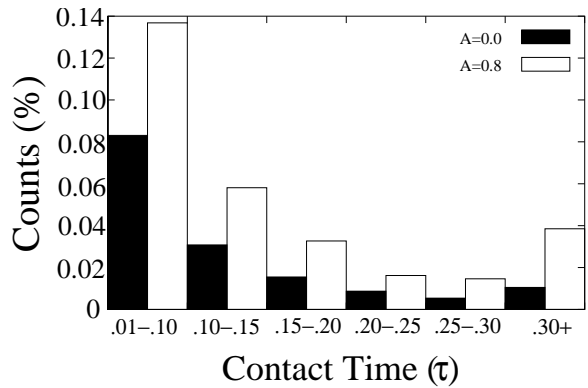}
\caption{ Contact time histogram within the flowing part of
the slab for cohesive energies $A=0.0$ (black) and $A=0.8$
(blue) with tilt angle $\theta=22^{o}$.  The shortest
contact time bin has been removed for clarity.}
\label{histogram}
\end{center}
\end{figure}

To more directly test that the breakdown of the Bagnold
constitutive relation is due to the growth of the number of
long-lived contacts in the flowing states with increasing
cohesive energy we compiled a contact time histogram. We
measured the times over which two particles remained in
contact allowing for both rolling and slipping at the
contact. Collecting this data for a few representative runs
having either no cohesion ($A=0$) or strong cohesion
($A=0.8$) we produced two contact time histograms as shown
in Fig. \ref{histogram}.  Both histograms reflect the
contact times in the horizontal slice of the slab between
$70\leq z\leq 100$ over a period of $10 \tau$. 
This vertical height was chosen so that it is entirely
contained within the flowing regime of the cohesive slab.
In fact, from our discussion of Fig.~\ref{bvsa}, we would
expect the largest population of long-lived contacts to be found in the upper part of the flowing region near the plug. For the cohesive system studied using the contact time histogram, we then expect the greatest density of long-time contacts to occur near   $z=200$. In both samples the most
common contact time is rather short:
$0<t_{\mbox{contact}}\leq 0.01 \tau$. This first bin has been
removed for clarity.  Comparing the two histograms we note
that although there are long-lived contacts in both the
cohesionless and the cohesive granular material, the number
of such contacts is dramatically increased in the cohesive
case. As discussed above the presence of long-lived
intergrain contacts in the cohesionless case is at least possibly due to the
nonzero compressibility of the particles. The enhancement
of  the number of such long-lived contacts, particularly
those having lifetimes greater than $\sim 0.3 \tau$,
with cohesion qualitatively supports our hypothesis that
such contacts are primarily responsible for the breakdown
of Bagnold scaling in the cohesive granular media.  The
relationship between this fraction of long-lived contacts
and a quantitative model for the parameter $\beta$ will be
explored in a future publication\cite{Brewster:05}.

\section{Summary}
\label{summary} We have studied the effects of cohesion in
steady-state 3D chute flow. Cohesive granular materials
generically form two different flow regimes: a solid region
(the plug) extending below the free surface characterized
by a vanishing shear rate of strain and a flowing region
below the plug. The width of this plug region is independent on total depth of the slab but dependent on the tilt angle, cohesive
energy, the mass density of the granular material and is
accounted for by postulating the appearance of a finite
yield stress in the material. The value of this yield
stress can be reasonably determined from estimates of the
mean cohesive force within the slab and the mean number of
such cohesive contacts.

We find that in the flowing state below the plug the
velocity profile is similar to the Bagnold velocity
profile: $v\propto z^{3/2}$. Nevertheless there remain
significant discrepancies between the observed flow profile
and the predicted Bagnold form.  In light of these
deviations from the Bagnold form, we suggest a modified
form of the Bagnold constitutive law that combines two
essentially independent methods of momentum transfer in the
flowing state. This modified Bagnold relation accounts for
the usual collisional momentum transfer between grains as
well as for momentum transfer due to the cohesive forces
acting through the long-term contacts between grains in the
flowing state.  This modified profile fits the data more
closely. Extracting from these fits to the data the ratio of
the stress transferred via each method, we find that this
ratio exhibits a few reasonable trends. The effect of the
long-term contacts grow with cohesive energy and within a single slab as
one approaches the plug.  This ratio
decreases with larger tilt angles; faster flowing slabs
should more efficiently break-up any long-lived particle
clusters.

To further test this hypothesis we directly measured a
contact time histogram within the flowing part of the slab.
A comparison between noncohesive and cohesive systems
indeed shows a larger number of long-time contacts in the
cohesive system. This result qualitatively supports our
hypothesis that these contacts are responsible for the
breakdown of the Bagnold constitutive law. To provide a
true quantitative model for the modified Bagnold constitutive
law that we propose it is necessary to calculate from a
more microscopic model the stresses transmitted by these
long-lived contacts. In order to do so we must determine
the spatial correlations between such contacts. These
contacts may, in fact, form spatially correlated clusters
or chains that span stream-lines in the flow or perhaps
represent randomly distributed pairs of particles (dimers)
that remain bound for mesoscopic periods during the flow.
Clearly, at a fixed density of such long-lived contacts the
stress transmission due to these contacts in the former
case would be significantly larger than in the latter
\cite{Brewster:05}.

Understanding the constitutive relation of cohesive
granular materials is clearly of fundamental importance to
the study of granular flows in a geophysical context as
well as to the handling of granular materials in industry
where small amounts of a wetting fluid create adhesion
contacts between the grains. Interestingly, this work
suggests that studying granular flows in the presence of
small amounts of cohesion allows one to break the expected
Bagnold scaling in a controllable and computationally
efficient manner. It is clear from this work and
others \cite{Silbert:01,Lois:05} that the non-zero
compliance of the grains leads to measurable deviations
from the Bagnold scaling; the Bagnold law holds in the limit
of hard spheres.  Studying significantly less
compliant grains, however, is computational difficult. The
further study of cohesive granular materials both
analytically and computationally should enable the
exploration of granular constitutive laws for physically
accessible systems that nearly, but not exactly obey the
Bagnold constitutive relation.

\appendix
\section{Derivation of the velocity from the Bagnold constitutive law}
\label{Bag-derivation} Bagnold scaling is derived from a
constitutive relation between the shear stress and the
strain rate
\begin{equation}
\label{bs} \sigma_{xz}=\kappa\dot{\gamma}^{2},
\end{equation}
where $\dot{\gamma}=\frac{\partial v_{x}(z)}{\partial z}.$
The steady-state Cauchy equation for $\sigma_{xz}$ is
\begin{equation}
\frac{\partial \sigma_{xz}}{\partial z}=\rho g\sin{\theta},
\end{equation}
therefore
\begin{equation}
\label{cauchy} \sigma_{xz}(z)=\rho g \sin{\theta}(H-z),
\end{equation}
where $H$ is the total height of the slab.  Solving for
$v_{x}(z)$ gives the velocity profile shown in
Eq.~\ref{Bag-vel} above.

Modified Bagnold scaling is derived from the same
constitutive relation used in deriving traditional Bagnold
scaling [Eq.~\ref{bs}] with the addition of a linear
term in $\dot{\gamma}$
\begin{equation}
\sigma_{xz}=\kappa\dot{\gamma}^{2}+\beta\dot{\gamma}.
\end{equation}
Solving this with the same Cauchy equation from
Eq.~(\ref{cauchy}) results in
\begin{equation}
v_{x}(z)=\frac{2}{3c\sqrt{\kappa}}\left[\left(G^{2}+cH\right)^{3/2}-\left(G^{2}+c(H-z)\right)^{3/2}\right]-Gz,
\end{equation}
where $c=\rho g\sin{\theta}$ and $G=\frac{\beta}{2\kappa}$


\begin{thebibliography}{99}
\bibitem{Pouliquen:02} O. Pouliquen and F. Chevoir, Physique {\bf 3}, 163 (2002).
\bibitem{Suzuki:71} A. Suzuki and T. Tanaka, Ind. Eng. Chem. Fundam. {\bf 10}, 84 (1971).
\bibitem{Augenstein:78} D. A. Augenstein and R. Hogg, Powd. Tech. {\bf 19}, 205 (1978).
\bibitem{Drake:90}T. J. Drake, J. Geophys. Res. {\bf 95}, 8681 (1990).
\bibitem{Ancey:96} C. Ancey, P. Coussot, and P. Evesque, Mech. Cohes-Fric. Matter {\bf 1}, 385 (1996).
\bibitem{Pouliquen:99} O. Pouliquen, Phys. Fluids {\bf 11}, 542 (1999).
\bibitem{Azanza:99} E. Azanza, F. Chevoir, and P. Moucheront, J. Fluid Mech. {\bf 400}, 199 (1999).
\bibitem{Ancey:02} C. Ancey, Phys. Rev. E {\bf 65}, 011304 (2002).
\bibitem{Pouliquen:04} O. Pouliquen, Phys. Rev. Lett. {\bf 93}, 248001 (2004).
\bibitem{Walton:93} O.R. Walton, Mech. Mater. {\bf 16}, 239 (1993).
\bibitem{Zheng:96} X.M. Zheng and J.M. Hill, Powd. Tech. {\bf 86}, 219 (1996).
\bibitem{Dippel:99} S. Dippel and D.E. Wolf, Comp. Phys. Comm. {\bf 121}, 284 (1999).
\bibitem{Hanes:00} D.M. Hanes and O.R. Walton, Powder Technol. {\bf 109}, 133 (2000).
\bibitem{Silbert:01} L.E.\ Silbert, D.\ Erta\c{s}, G.S.\ Grest, T.C.\ Halsey, D. Levine, and S.J. Plimpton, Phys. Rev. E {\bf 64}, 051302 (2001).
\bibitem{Silbert:02}
L.E. Silbert, G.S. Grest, S.J. Plimpton, and D. Levine,
Phys. Fluids {\bf 14}, 2637 (2002).
\bibitem{Silbert:03}
L.E. Silbert, J.W. Landry, and G.S. Grest, Phys. Fluids {\bf
15}, 1 (2003).
\bibitem{Haff:83} P.K. Haff, J. Fluid Mech. {\bf 134}, 401 (1983).
\bibitem{Savage:83} S.B. Savage, {\em Advances in Micromechanics of Granular Materials} eds. J.T. Jenkins and M. Satake (Elsevier, New York) (1992).
\bibitem{Savage:98} S.B. Savage, J. Fluid Mech. {\bf 377}, 1 (1998).
\bibitem{Jenkins:83} J.T. Jenkins and S.B. Savage J. Fluid Mech. {\bf 130}, 187 (1983).
\bibitem{Losert:00}
W. Losert, L. Bocquet, T.C. Lubensky, and J.P. Gollub, Phys. Rev. Lett. {\bf 85}, 1428 (2000).
\bibitem{Bocquet:01} L. Bocquet, W. Losert, D. Schalk, T.C. Lubensky, and J.P. Gollub, Phys. Rev. E {\bf 65}, 011307 (2001).
\bibitem{Bocquet:02}
L. Bocquet, J. Errami, and T.C. Lubensky, Phys. Rev. Lett. {\bf 89}, 184301 (2002).
\bibitem{Mills:00} P. Mills, D. Loggia, and M. Tixier, Eur. Phys. J. E {\bf 1}, 5 (2000).
\bibitem{Louge:03} M.Y. Louge, Phys. Rev. E {\bf 67}, 061303 (2003).
%
\bibitem{Bazant:05} M.Z. Bazant, unpublished (2005).  
\bibitem{Bagnold:54} R.A. Bagnold, Proc. R. Soc. London, Ser A {\bf 225}, 49 (1954); 
{\bf 249}, 235 (1956).
\bibitem{Cundall:79} P.A. Cundall and O.D.L. Strack, Geotechnique {\bf 29}, 47 (1979).
\bibitem{Ertas:02} D. Erta\c{s} and T.C. Halsey, Europhys. Lett. {\bf 60}, 931 (2002);
T.C. Halsey and D. Erta\c{s}, cond-mat/0506170 (2005).
\bibitem{Lois:05} G. Lois, A. Lemaitre, and J.M. Carlson, cond-mat/0501535 (2005).
\bibitem{Brewster:05} R.\ Brewster and A.J.\ Levine, in preparation (2005).

\end{thebibliography}
\end{document}